\documentclass{sig-alternate}
\usepackage[ruled,vlined]{algorithm2e}
\usepackage{graphicx}
\usepackage{url}
\usepackage{paralist}
\usepackage{tabularx}
\usepackage{balance}
\usepackage{multirow}
\usepackage{multicol}
\usepackage{pbox}
\usepackage{mathtools}
\usepackage[TABBOTCAP]{subfigure}
\usepackage{algorithmic}
\usepackage{graphicx}
\usepackage{paralist}
\usepackage{tabularx}
\usepackage{url}
\usepackage{balance}
\usepackage{multirow}
\usepackage{multicol}
\usepackage{amsmath}
\usepackage{setspace}
\usepackage[bookmarks=false]{hyperref}
\usepackage{verbatim}
\usepackage{float}
\usepackage[normalem]{ulem}
\usepackage{xspace}
\usepackage[usenames]{color}
\usepackage{colortbl}
\usepackage{amssymb}
\usepackage{ifthen}
\usepackage{pifont}
\usepackage{tikz}
\newcommand*\circled[1]{\tikz[baseline=(char.base)]{
            \node[shape=circle,draw,inner sep=0.5pt] (char) {#1};}}

\include{macro}

\newcommand{\Fusionsp}{{\sc Fusion~}}
\newcommand{\Fusion}{{\sc Fusion}}
\newcommand{\Fusions}{{\sc Fusion's~}}

\newcommand\blfootnote[1]{%
  \begingroup
  \renewcommand\thefootnote{}\footnote{#1}%
  \addtocounter{footnote}{-1}%
  \endgroup
}

\begin{document}
\CopyrightYear{2016} 
\setcopyright{acmlicensed}
\conferenceinfo{ICSE '16 Companion,}{May 14 - 22, 2016, Austin, TX, USA}
\isbn{978-1-4503-4205-6/16/05}\acmPrice{\$15.00}
\doi{http://dx.doi.org/10.1145/2889160.2889177}

\title{FUSION: A Tool for Facilitating and Augmenting Android Bug Reporting}

\numberofauthors{1} \author{
\alignauthor
Kevin Moran, Mario Linares-V\'asquez, Carlos Bernal-C\'ardenas, Denys Poshyvanyk\\
\affaddr{College of William \& Mary}\\
\affaddr{Department of Computer Science}\\
\affaddr{Williamsburg, VA 23187-8795, USA}\\
\email{\{kpmoran, mlinarev, cebernal, denys\}@cs.wm.edu}
}

\maketitle
\begin{abstract}
As the popularity of mobile smart devices continues to climb the complexity of ``apps" continues to increase, making the development and maintenance process challenging.  Current bug tracking systems lack key features to effectively support construction of reports with actionable information that directly lead to a bug's resolution. In this demo we present the implementation of a novel  bug reporting system, called \Fusion, that facilitates users including reproduction steps in bug reports for mobile apps. \Fusionsp links user-provided information to program artifacts extracted through static and dynamic analysis performed before testing or release.  Results of preliminary studies demonstrate that \Fusionsp both effectively facilitates reporting and allows for more reliable reproduction of bugs from reports compared to traditional issue tracking systems by presenting more detailed contextual app information. Tool website: \url{www.fusion-android.com} Video url: \url{https://youtu.be/AND9h0ElxRg}
\end{abstract}

\vspace{-0.3cm}

\ccsdesc[500]{Software and its engineering~Software maintenance tools}
\ccsdesc[500]{Software and its engineering~Software testing and debugging}
\ccsdesc[500]{Software and its engineering~Maintaining software}
\ccsdesc[300]{Software and its engineering~Software evolution}

\printccsdesc

\vspace{-0.3cm}

\section{Introduction}
\label{sec:intro}

	It is clear that as smart device usage reaches ubiquitous levels (e.g., 2.7 billion active smartphone users in 2014\cite{24MobilityReport}), developers need tools to support them in maintaining high-quality apps.  Software maintenance activities are known to be generally expensive and challenging \cite{25Tassey:NIST} and one of the most important maintenance tasks is bug report resolution.  However, current bug tracking systems such as Bugzilla \cite{bugzilla}, Mantis \cite{mantis}, the Google Code Issue Tracker \cite{google-code}, the GitHub Issue Tracker \cite{github-it}, and commercial solutions such as JIRA \cite{jira} rely mostly on unstructured natural language bug descriptions.  While these descriptions can be supplemented with structured information such as reproduction steps or stack traces, and files such as screenshots, the inclusion of such information typically depends on the reporter's experience
\\
\\
\\		
and attitude towards providing these details. Previous studies have also shown that the information most useful to developers is often the most difficult for reporters to provide and that the lack of this information is a major reason behind non-reproducible bug reports \cite{4Joorabchi:MSR14, 3Bettenburg:FSE08}. Therefore, the reporting process can be cumbersome, and the additional effort means that many users are unlikely to enhance their reports with extra information \cite{11Bettenburg:MSR08,31Davies:ESEM2014,32Bettenburg:ICSM08, 34Aranda:ICSE09}.\blfootnote{This work is supported in part by the NSF CCF-1218129 and CCF-1525902 grants. Any opinions, findings, and conclusions expressed herein are the authors' and do not necessarily reflect those of the sponsors.}

	The above issues point to a more prominent problem for bug tracking systems in general: the \textit{lexical gap} that normally exists between bug reporters (e.g., testers, beta users) and developers.  Reporters typically only have functional knowledge of an app, even if they have development experience themselves, whereas the developers working on an app tend to have intimate code level knowledge.  When a developer reads and attempts to comprehend (or reproduce) a bug report, she has to bridge this gap, reasoning about the code level problems from the high-level functional description in the bug report.  If the lexical gap is too wide the developer may not be able to reproduce and/or subsequently resolve the bug report.  	

	To address this fundamental problem of making bug reports more useful (and reproducible) for mobile applications, this paper presents the implementation of a novel tool, called \Fusion, that facilitates reporters creating detailed bug reports in order to provide more actionable information to developers.  \Fusionsp implements the novel approach that was presented and evaluated in our previous work \cite{Moran:FSE2015}.  \Fusionsp first employs fully automated static and dynamic analysis techniques to gather screen-shots and other relevant information about an app before it is released for testing. Reporters then interact with the web-based report generator using the auto-completion features in order to provide the bug reproduction steps.  By linking the information provided by the user with features extracted through static and dynamic analyses, \Fusionsp presents an augmented bug report to the developer that contains actionable information with well-defined steps to reproduce a bug. 		

\section{The FUSION Reporting Tool}
\label{sec:approach}
    
\Fusions current target user base consists of two major groups: \textit{mobile application developers} and \textit{beta} users or \textit{testers}.  As such there are two user-facing scenarios for the tool's operation.  From a developer's perspective the workflow is purposefully simple; the only action required is the submission of an \texttt{.apk} of the latest build of the application for which they want to enable bug reporting.  Once this file is submitted, \Fusions automated program analysis techniques extract the information necessary to facilitate process of creating a bug from the reporters end.  Once reports are filled out, developers can access them through a web application.  From a reporter's perspective, they construct a report using a web interface.  After selecting the application for which they would like to report a bug, the reporter enters a brief description and contextual information about the bug (e.g. device used, screen orientation).  They then use a series of auto-filled combo boxes to construct reproduction steps, including screenshots.

\subsection{FUSION Architecture}
 
Fusion's architecture can be seen in Figure \ref{Design}. First, \Fusionsp collects information related to the GUI components and event flow of an app using the \textit{Static Application Analyzer} and the \textit{Dynamic Analysis Engine}.  Then the tool leverages the information collected during the analysis to facilitate a reporter constructing a detailed bug report with reproduction steps, and screenshots.
    The static and dynamic analyses must be performed before each version of an app is released for testing or before it is published to end users.  Both program analysis components store their extracted data in the \Fusionsp database (Fig. \ref{Design} - \circled{3}).
    
\subsubsection{Static Analysis}
    
    The goal of the \emph{Static App Analyzer} (Fig. \ref{Design} - \circled{1})
   is to extract all of the GUI components and associated information from the app source code.   For each GUI component, it extracts: (i) possible actions on that component, (ii) type of the component (e.g., Button, Spinner), (iii) activities the component is contained within, and (iv) class files where the component is instantiated.  Thus, this results in a universe of possible components within the domain of the application, and establishes traceability links connecting GUI-components to code specific information such as the class or \texttt{activity} where they are located.  
    
    The \emph{Static App Analyzer} utilizes several tools to extract the information outlined above. First it uses the  {\tt dex2jar}\cite{dex2jar} and  {\tt jd-cmd} \cite{jd-cmd}  tools for decompilation; then, it converts the source files to an XML-based representation using {\tt srcML} \cite{srcml}. We also use {\tt apktool} \cite{apktool} to extract the resource files from the app's APK.  The {\tt id}s and types of GUI components were extracted from the xml files located in the app's resource folders (\textit{i.e.},  {\tt /res/layout} and  {\tt /res/menu} of the decompiled application or src).  Using the {\tt srcML} representation of the source code, we are able to parse and link the GUI-component information to extracted app source files.  
    
\subsubsection{Dynamic Analysis}

\begin{figure}[tb]
\centering
\includegraphics[width=\linewidth]{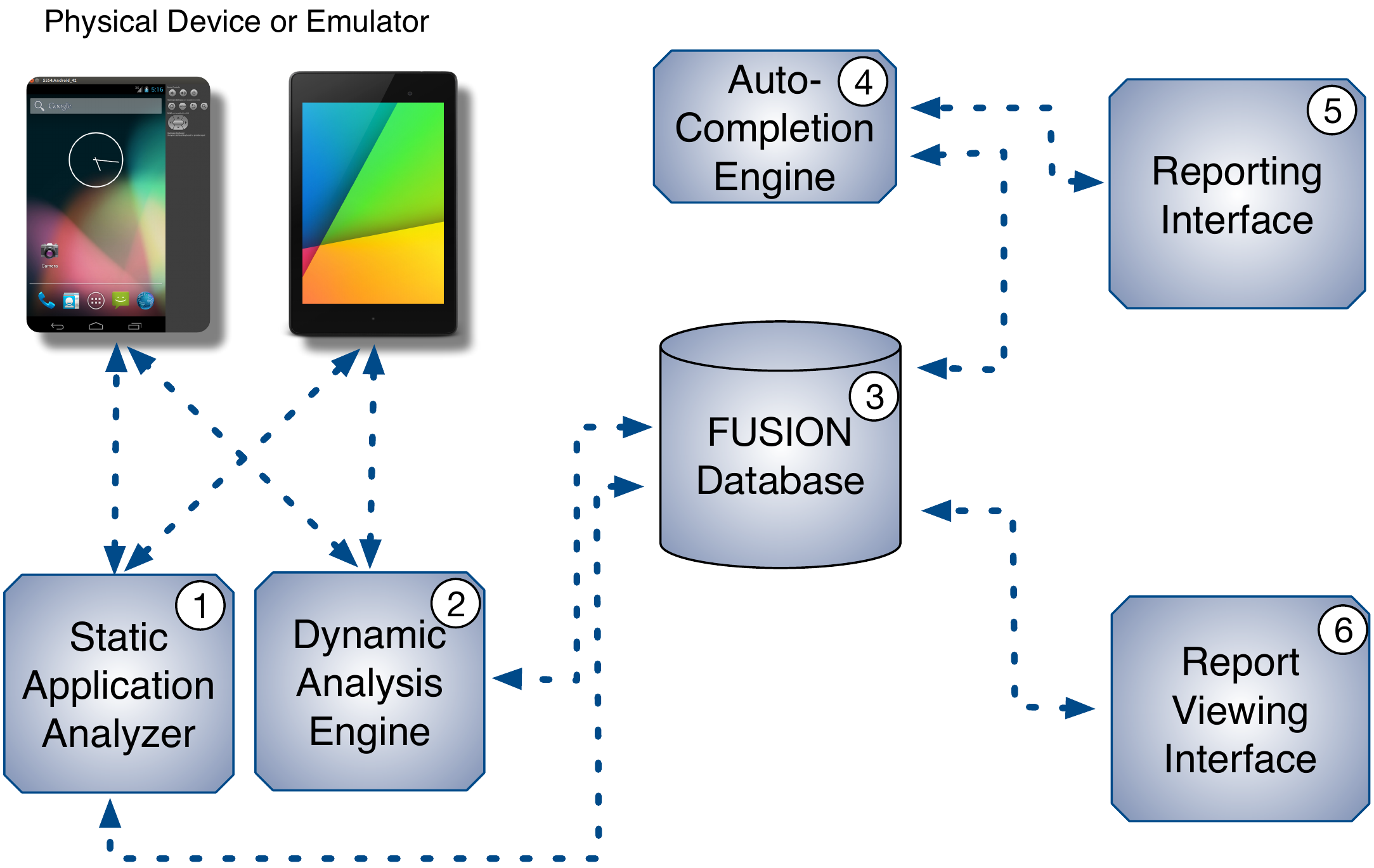}
\vspace{-0.7cm}
\caption{\Fusionsp Architecture}
\label{Design}
\end{figure}

The \emph{Dynamic Analysis Engine} (Fig. \ref{Design} - \circled{2}) is used to glean dynamic contextual information and enhance the data- base with both run-time GUI and application event-flow information. The goal of the \emph{Engine} is to explore an app in a systematic manner, ripping and extracting run-time information related to the GUI components during execution including: (i) the text associated with different GUI components (e.g., the ``Send'' text on a button to send an email message), (ii) whether the GUI component triggers a transition to a different activity, (iii) the action performed on the GUI component during systematic execution, (iv) full screen-shots before and after each action is performed, (v) the location of the GUI component object on the test device's screen, (vi) the current activity and window of each step, (vii) screen-shots of the specific GUI component, and (viii) the object index of the GUI component (to allow for differentiation between different instantiations of the same GUI component on one screen). 

    The \emph{Engine} performs this systematic exploration of the app using the {\tt UIAutomator} \cite{uiautomator} framework included in the Android SDK.  This systematic execution of the app is similar to existing approaches in GUI ripping \cite{Amalfitano:ASE2012,Azim:OOPSLA2013,Machiry:FSE2013,Linares:MSR15}. Using the {\tt UIAutomator} framework allows us to capture cases that are not captured in previous tools such as pop-up menus that exist within menus, internal windows, and the onscreen keyboard.  To effectively explore the application we implemented our own version of a systematic depth-first search (DFS) algorithm for application traversal that performs click events on all the clickable components in the GUI hierarchy reachable using the DFS heuristic.
      
\subsection{Reporting Bugs with FUSION}  
       
                During the \emph{Report Generation Phase}, \Fusionsp aids the reporter in constructing the steps needed to recreate a bug by making suggestions based upon the ``potential" GUI state reached by the declared steps.  This means for each step $s$, \Fusionsp infers --- online --- the GUI state $GUI_s$ in which the target app should be by taking into account the history of steps.   For each step, \Fusionsp verifies that the suggestion made to the reporter is correct by presenting the reporter with contextually relevant screen-shots, where the reporter selects the screen-shot corresponding to the current action she wants to describe. It should be noted that the current scope of \Fusionsp is limited to \textit{functional, gui-based bugs} however, we have plans to extend this in future work.

\subsubsection{Report Generator User Interface}

\begin{figure}[t]
\centering
\includegraphics[width=\linewidth]{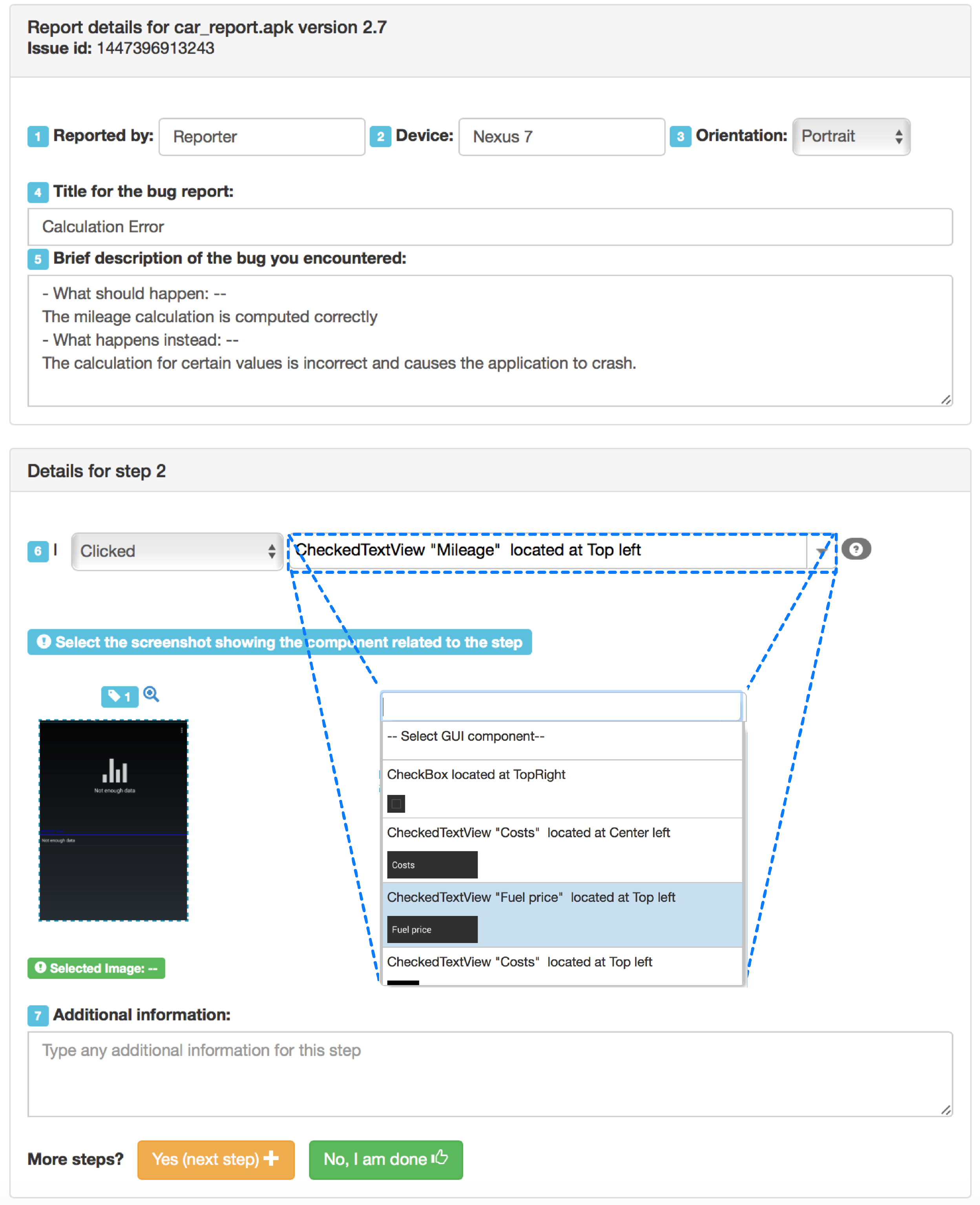}
\vspace{-0.5cm}
\caption{\Fusionsp Reporting Interface}
\label{front_end}
\end{figure}

    After first selecting the app to report an issue for, a reporter interacts with \Fusionsp by filling in some identifying information (i.e., name, device, title) and a brief textual description of the bug in the top half of the UI.  Next, the reporter inputs the steps to reproduce the bug using the auto-completion boxes in a step-wise manner, starting from the initial screen of a cold app launch\footnote{Cold-start means the first step is executed on the first window and screen displayed directly after the app is launched.}, and proceeds until the list of steps to reproduce the bug is exhausted.  According to the various fields in Figure \ref{front_end}, the reporter would first fill in their (i) \textit{name} (Field 1), (ii) \textit{device} (Field 2), (iii) \textit{screen orientation} (Field 3), (iv) a \textit{bug report title} (Field 4), and (v) a \textit{brief description of the bug} (Field 5). 
     
     The first drop down list (see Figure \ref{front_end} - Field 6) corresponds to the possible actions a user can perform at a given point in app execution.  For example, let's say the reporter selects \textit{click} as the first action in the sequence of steps as shown in Figure \ref{front_end}. The possible actions considered in \Fusionsp  are \textit{click(tap), long-click(long-touch), type}, and \textit{swipe}.  The \textit{type} action corresponds to a user entering information from the device keyboard.  When the reporter selects the \textit{type} option, we also present them with a text box to collect the information she typed in the Android app.

\begin{figure}[t]
\centering
\includegraphics[width=\linewidth]{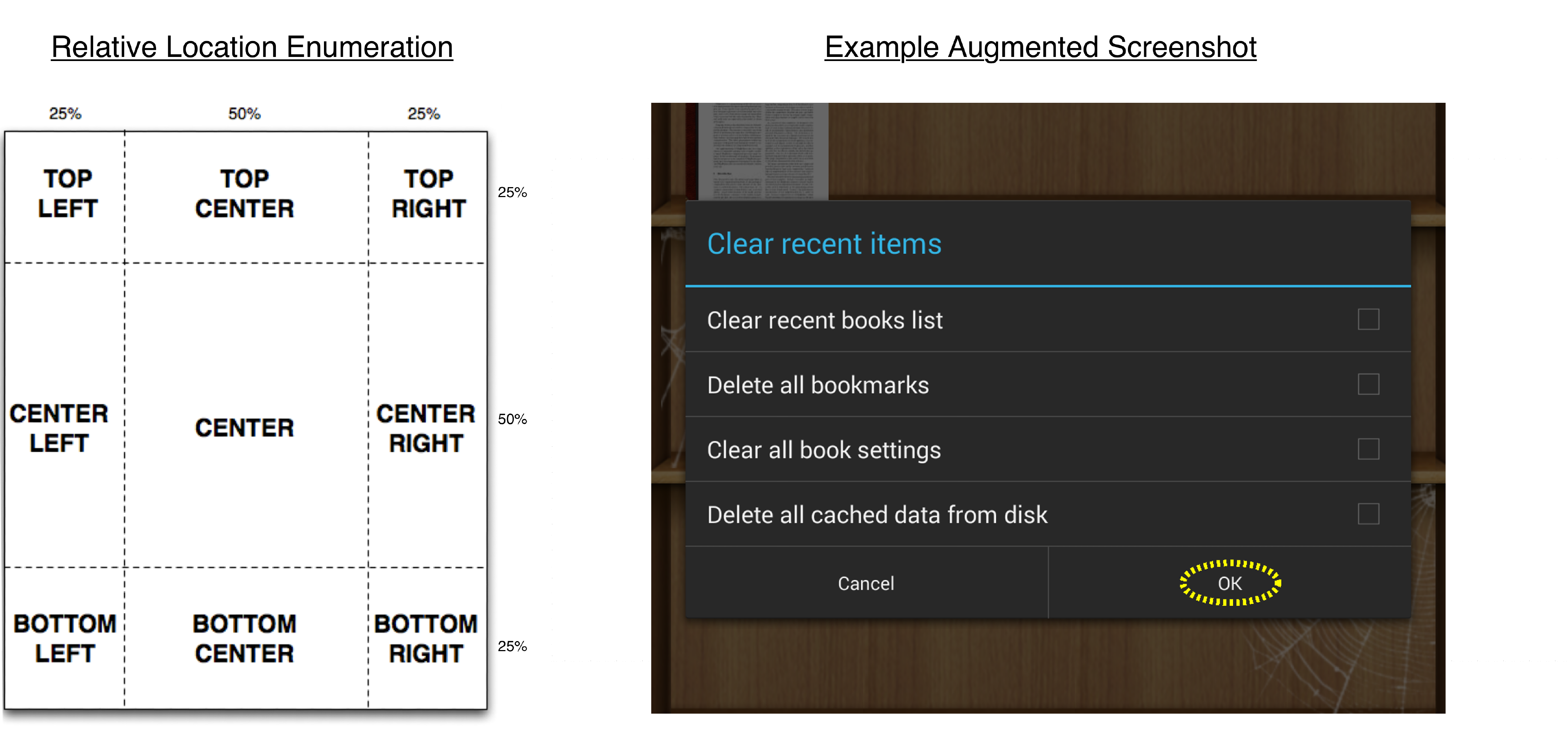}
\vspace{-0.8cm}
\caption{Relative Location Enumeration and Example Augmented Screenshot}
\label{example_screenshot}
\end{figure}

	The second dropdown list (see Figure \ref{front_end} - outlined in blue) corresponds to the component associated with the action in the step. \Fusionsp presents the following information, which can also be seen in Figure \ref{front_end}: (i) the \textit{type of component} that is being operated upon, (e.g. button, spinner, checkbox),  (ii)  the \textit{text} associated with or located on the component, (iii) the \textit{relative location} of the component on the screen according to the parameters in Figure \ref{example_screenshot}, and (iv) : an in-situ (i.e., embedded in the dropdown list) \textit{image} of the instance of the component.  The relative location is displayed here to make it easier for reporters to reason about the on-screen location, rather than reasoning about pixel values.  In the example from Figure \ref{front_end} above, \Fusionsp will populate the component dropdown list with all of the clickable components in the \texttt{main} Activity since this is the first step and the selected action was \textit{click}.  
	    
   \Fusionsp uses two techniques to handle instances of seemingly identical GUI-components appearing on the same screen. \textit{First}, it differentiates each duplicate component in the list through specifying text ``Option \#''.  \textit{Second} \Fusionsp attempts to confirm the component entered by the reporter at each step by fetching screen-shots from the \Fusionsp database representing the entire device screen (e.g., Figure \ref{example_screenshot}).
    
     After the reporter makes selections from the drop-down lists, they have an opportunity to enter additional information for each step (e.g., a button had an unexpected behavior) in a natural language text entry field.  For instance, the reporter might indicate that after pressing the ``OK" button the pop-up window took longer than expected to disappear.  
   
   \subsubsection{Report Generator Auto-Completion Engine}
 
 The \emph{Auto-Completion Engine} of the web-based report generator (Figure \ref{Design}-\circled{4}) uses the information collected up-front during the \textit{Analysis Phase}. When \Fusionsp suggests completions for the drop-down menus, it queries the database for the corresponding state of the app event flow and suggests information based on the past steps that the reporter has entered.  Since we always assume a ``cold'' application start, the \emph{Auto-Completion Engine} starts the reproduction steps entry process from the app's \texttt{main} Activity.  We then track the reporter's progress through the app using predictive measures based on past steps.    
	
	\Fusionsp presents components to the reporter at the granularity of \texttt{activities}, or application screens.  During the suggestion process, \Fusionsp looks back through the history of reported user interactions and looks for possible transitions from the previous steps to future steps depending on the history of the components interacted with.  If \Fusionsp is unable to capture the last few steps from the reporter due to the incomplete application execution model mentioned earlier, then \Fusionsp presents the possibilities from all known screens of the application.  Due to the limited nature of the DFS heuristic used by the \textit{Dynamic Analysis Engine}, there may be action-Gui-Component pairs that are not available in the auto-filled combo boxes.  To handle these cases gracefully, we allow the reporter to select a special option when they cannot find a component in the auto-completed combo box.  When picking this option, the reporter would manually fill in (i) the type of the component, (ii) any text associated with the GUI-component, and (iii) the relative location of the GUI-component (according to the screen regions listed in Figure \ref{example_screenshot}).
		 
   \begin{figure}[t]
\centering
\includegraphics[width=0.95\linewidth]{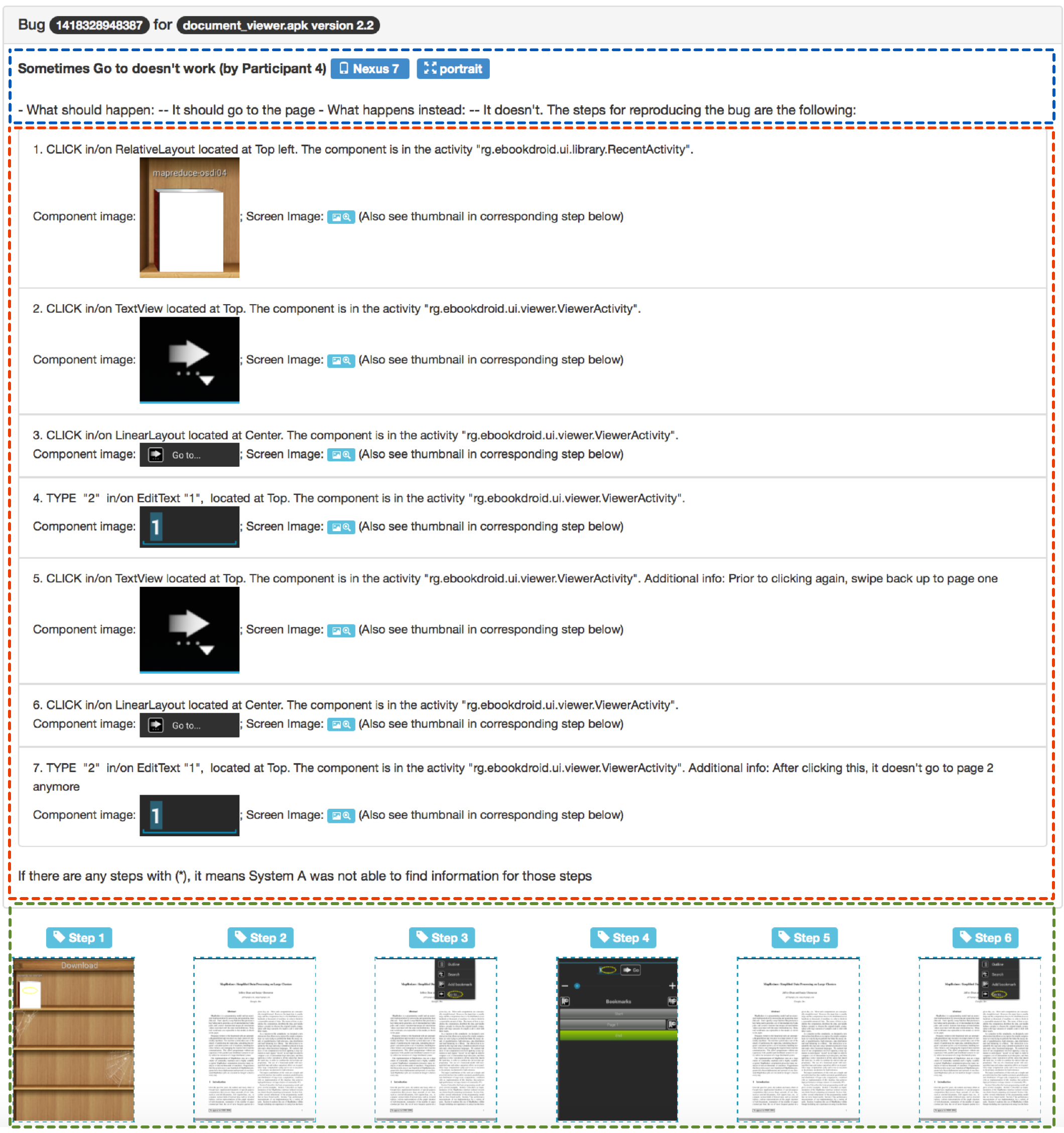}
\vspace{-0.4cm}
\caption{Example \Fusionsp Bug Report}
\label{report}
\end{figure}  

\subsubsection{Viewing FUSION Reports}

      The \emph{Auto Completion Engine} saves each step to the database as reporters complete bug reports.  Once a reporter finishes filling out the steps and completes the data entry process, a screen containing the final report, with an automatically assigned unique ID, is presented to the reporter and saved to the database for a developer to view later (see Figure \ref{report} for an example report). 

       The Report presents information to developers in three major sections: First, preliminary information including the report title, device, and short description (shown in Figure \ref{report} in blue).  Second, a list of the Steps with the following information regarding each step is displayed (highlighted in blue in Figure \ref{report}): (i) The action for each step, (ii) the type of a component, (iii) the relative location of the component, (iv) the \texttt{activity} Java class where the component is instantiated in the source code, and (v) the component specific screenshot.  Third, a list of full screen-shots corresponding to each step is presented at the bottom of the page, thus, the developer can trace the steps through each application screen (this section is highlighted in green in Figure \ref{report}).

\section{Evaluation}
\label{sec:evaluation}
To evaluate \Fusionsp (see full details in \cite{Moran:FSE2015}) we investigated its ease of use, as well as the reproducibility of the \Fusionsp reports compared to reports created using Google Code Issue Tracker (GCIT).  First, in a \textit{bug-creation study} we recruited eight students (four undergraduate or \textit{non-experts} and four graduate or \textit{experts}) to construct bug reports using \Fusionsp and GCIT --- as a representative of traditional bug tracking systems--- for 15 real-world world bugs for 14 open-source apps from F-Droid \cite{fdroid}.  We collected survey responses from these participants regarding the ease of use and user preferences of each tool.  Next, in a \textit{bug-reproduction study} we evaluated the reproducibility of the \Fusionsp and GCIT reports generated by the first group of participants. These reports (120 for each type) and the original bug reports extracted from the respective app issue trackers were evaluated by a new set of 20 graduate student participants through attempted bug reproduction on physical devices.  \textbf{The results of this study indicate that  developers using \Fusionsp reports are able to reproduce more reports (107 out of 120) compared to traditional bug tracking systems such as the GCIT (97 of 120).}  
\bigskip

\section{Demo Remarks and Future Work}
\label{sec:concl}
\vspace{-0.1cm}
In this demo, we presented \Fusion, a novel implementation of an enhanced bug reporting system for Android applications.  \Fusionsp facilitates reporters crafting detailed reports containing reproduction steps, screenshots, and traceability links to code artifacts, by informing the reporting process with data gleaned from static and dynamic program analyses.  As future work, we plan to (i) investigate more sophisticated methods of modeling program behavior, such as using statistical language models~\cite{Linares:MSR15}, (ii)  to improve our DFS engine through supporting more gestures, (iii) and to use FUSION as a tool for reporting feature requests.

\vspace{-0.2cm}
\balance
\small{
\bibliography{ms}

\begin{thebibliography}{10}

\bibitem{uiautomator}
Android uiautomator
  \url{http://developer.android.com/tools/help/uiautomator/index.html}.

\bibitem{apktool}
apktool \url{https://code.google.com/p/android-apktool/}.

\bibitem{bugzilla}
Bugzilla issue tracker \url{https://bugzilla.mozilla.org}.

\bibitem{dex2jar}
dex2jar \url{https://code.google.com/p/dex2jar/}.

\bibitem{fdroid}
F-droid. \url{https://f-droid.org/}.

\bibitem{github-it}
Github issue tracker \url{https://github.com/features}.

\bibitem{google-code}
Google code issue tracker
  \url{https://code.google.com/p/support/wiki/IssueTracker}.

\bibitem{jd-cmd}
jd-cmd decompiler \url{https://github.com/kwart/jd-cmd}.

\bibitem{jira}
Jira bug reporting system \url{https://www.atlassian.com/software/jira}.

\bibitem{mantis}
Mantis bug reporting system \url{https://www.mantisbt.org}.

\bibitem{srcml}
srcml \url{http://www.srcml.org}.

\bibitem{Amalfitano:ASE2012}
D.~Amalfitano, A.~R. Fasolino, P.~Tramontana, S.~De~Carmine, and A.~M. Memon.
\newblock Using gui ripping for automated testing of android applications.
\newblock ASE'12, pages 258--261.

\bibitem{34Aranda:ICSE09}
J.~Aranda and G.~Venolia.
\newblock The secret life of bugs: Going past the errors and omissions in
  software repositories.
\newblock In {\em ICSE'09}, pages 298--308, 2009.

\bibitem{Azim:OOPSLA2013}
T.~Azim and I.~Neamtiu.
\newblock Targeted and depth-first exploration for systematic testing of
  android apps.
\newblock In {\em OOPSLA'13}, pages 641--660, 2013.

\bibitem{3Bettenburg:FSE08}
N.~Bettenburg, S.~Just, A.~Schr\"{o}ter, C.~Weiss, R.~Premraj, and
  T.~Zimmermann.
\newblock What makes a good bug report?
\newblock In {\em FSE'08}, pages 308--318, 2008.

\bibitem{32Bettenburg:ICSM08}
N.~Bettenburg, R.~Premraj, T.~Zimmermann, and S.~Kim.
\newblock Duplicate bug reports considered harmful... really?
\newblock In {\em ICSM'08}, pages 337--345, 2008.

\bibitem{11Bettenburg:MSR08}
N.~Bettenburg, R.~Premraj, T.~Zimmermann, and S.~Kim.
\newblock Extracting structural information from bug reports.
\newblock In {\em MSR'08}, pages 27--30, 2008.

\bibitem{31Davies:ESEM2014}
S.~Davies and M.~Roper.
\newblock What's in a bug report?
\newblock ESEM '14, pages 26:1--26:10.

\bibitem{4Joorabchi:MSR14}
M.~Erfani~Joorabchi, M.~Mirzaaghaei, and A.~Mesbah.
\newblock Works for me! characterizing non-reproducible bug reports.
\newblock In {\em MSR'14}, pages 62--71, 2014.

\bibitem{24MobilityReport}
Ericsson.
\newblock Ericsson mobility report.
\newblock
  http://www.ericsson.com/res/docs/2014/ericsson-mobility-report-november-2014.pdf.

\bibitem{Linares:MSR15}
M.~Linares-V\'{a}squez, M.~White, C.~Bernal-C\'{a}rdenas, K.~Moran, and
  D.~Poshyvanyk.
\newblock Mining android app usages for generating actionable gui-based
  execution scenarios.
\newblock In {\em MSR'15}, pages 111--122, 2015.

\bibitem{Machiry:FSE2013}
A.~Machiry, R.~Tahiliani, and M.~Naik.
\newblock Dynodroid: An input generation system for android apps.
\newblock ESEC/FSE'13, pages 224--234.

\bibitem{Moran:FSE2015}
K.~Moran, M.~Linares-V\'{a}squez, C.~Bernal-C\'{a}rdenas, and D.~Poshyvanyk.
\newblock Auto-completing bug reports for android applications.
\newblock In {\em ESEC/FSE'15}, pages 673--686, 2015.

\bibitem{25Tassey:NIST}
G.~Tassey.
\newblock The economic impacts of inadequate infrastructure for software
  testing.
\newblock Technical report, National Institute of Standards and Technology,
  2002.

\end{thebibliography}
\bibliographystyle{abbrv}}

\balancecolumns

\end{document}